\documentclass[twocolumn,preprintnumbers,amsmath,amssymb]{revtex4}
\usepackage{graphicx}% Include figure files
\usepackage{dcolumn}% Align table columns on decimal point
\usepackage{bm}% bold math

\begin{document}
%\pagestyle{empty}

%\preprint{APS/123-QED}

\title{HIGH-$p_T$ SPECTRA OF CHARGED HADRONS  \\
IN Au+Au COLLISIONS AT $\sqrt {s_{NN}} =9.2$~GeV IN STAR\\
}

\author{M.V.~Tokarev  for the STAR Collaboration\footnote{Speaker}\footnote{E-mail: tokarev@sunhe.jinr.ru}
}
\affiliation{Joint Institute for Nuclear Research, Dubna, Russia}
%\date{\today}
\begin{abstract}
The production of hadrons in heavy ions collisions at high $p_T$ provides
an important information on mechanism of particle formation and
constituent  energy loss in medium. Such information is needed for search
of a Critical Point and signatures of phase transition. Measurements by
the STAR Collaboration of charged hadron production in Au+Au collisions at
$\sqrt {s_{NN}}$=9.2~GeV over a wide transverse momentum $p_T=0.2-4.$~GeV/c
and at mid-rapidity range are reported.
It allows for a first measurement of
the spectra for charged hadrons at high  $p_T$ at this energy.
The spectra demonstrate
the dependence on centrality which enhances with $p_T$. The constituent
energy loss and its dependence on transverse momentum of particle, and
centrality of collisions are estimated in the $z$-scaling approach.
\\[5mm]
PCAS numbers: 25.75.-q
\vskip 5mm
%"Physics of Fundamental Interactins", Session-conference of Russian Academy of %Sciences,
%November 23-27, 2009,
%Institute of Theoretical and Experimental Physics, Moscow, Russia
\end{abstract}

%\pacs{Valid PACS appear}
%Use showkeys class option if keyword
\maketitle

%\section{\label{sec:level1}Introduction}

\section{Introduction}

Heavy ion collisions at RHIC have provided evidence that a
new state of nuclear matter exists \cite{WhitePaper}.
This new state is characterized by a suppression
of particle production
at high $p_T$ \cite{suppression}, a large amount
of elliptic flow $(v_2)$, constituent quarks
(NCQ) scaling of $v_2$ at intermediate $p_T$
\cite{v2flow} and enhanced correlated yields
at large $\Delta \eta $ and $\Delta \phi \simeq 0$ (ridge) \cite{ridge}.

To understand the properties of the system in the framework of
the Quantum Chromodynamics (QCD)  is one of the main goals
of high energy heavy ion collision experiments at RHIC and SPS.

Calculations in lattice QCD (see \cite{Stepanov,Karsch} and references therein)
 indicate that the energy density  ($3-5$~GeV/fm$^3$) and temperature
 ($T\simeq 170$~K) reached in central Au+Au collisions at RHIC
are enough  to observe  signatures
(enhancement of multiplicity, transverse momentum and
particle ratios fluctuations,
long-rang correlations, strange-hadron abundances,...)
of a possible phase transition from hadronic
to  quark and gluon degrees of freedom.
Nevertheless, a clear  indication of such a transition has yet to be observed.
This has been widely discussion in the literature
\cite{Mitchell,Sorensen,Mohanty,Braun}.
The principal challenge remains localization of a Critical
Point on the QCD phase diagram.
Near the QCD Critical Point,
several thermodynamic properties of the system
such as the heat capacity, compressibility,
correlation length
are expected to diverge with a power-law behavior in the
variable $\epsilon  = (T - T_c)/T_c$,
where $T_c$ is the critical temperature.
The rate of the
divergence can be described by a set of critical exponents.
 The  critical exponents
are universal in the sense that they depend only on  degrees of freedom
in the theory and their symmetry, but not on other details of the interactions.
This scaling postulate is the central concept of the theory
of critical phenomena \cite{Stanley}.

An important step towards understanding the structure of the QCD phase diagram
is  systematic analysis of particle production as a function of collision energy, centrality and collisions species.
Assuming the system is thermalized, temperature $T$
and baryon chemical potential $\mu_B$ can be determined.
A search for the location of a possible Critical Point on
the $\{ {T_c,\mu_c} \}$ phase diagram, can be done by varying
the beam energy.
The proposed Beam Energy Scan (BES) has been tasked to
carry out this search \cite{BES}.

A first test run for Au+Au collisions at $\sqrt {s_{NN}} = 9.2$~GeV made
by the RHIC has allowed the STAR Collaboration to obtain the first
results on
identified particle $(\pi^{\pm}, K^{\pm}, p,\bar p)$ production,
azimuthal anisotropy,  interferometry measurements \cite{AuAu9GeV},
and on high-$p_T$ spectra of charged hadron production, which are
reported in this paper.

\section{Experiment and data analysis}

\begin{figure}
\hspace*{0mm}
\includegraphics[width=91mm,height=55mm]{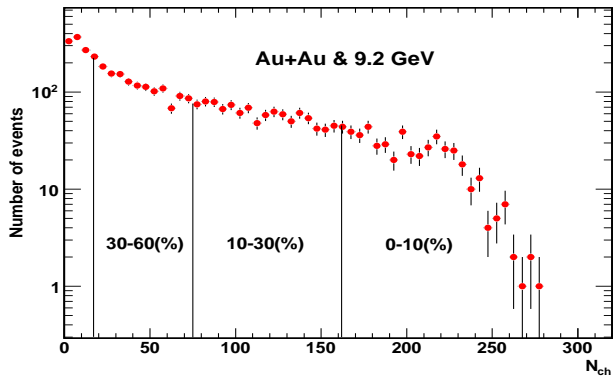}
\caption{
Uncorrected charged particle multiplicity distribution measured in the TPC within
$\eta <$~0.5 in Au + Au collisions at $\sqrt s_{NN}= 9.2$~GeV.
The vertical lines reflect the
centrality selection criteria
(centrality classes - 0--10\%, 10--30\%, 30--60\%) used in the paper \cite{AuAu9GeV}.
Errors are statistical only.}
\end{figure}

The data presented here are from Au+Au collisions
at $\sqrt {s_{NN}} = 9.2$~GeV,
recorded by the STAR experiment
in  a short run conducted in 2008 at RHIC.
The energy of collided Au ions is less than
the injection energy.
The data taking period
covered about five hours.
There are  $\simeq 4000$ good events collected at about 0.6 Hz
 which are used for this analysis.
The main detector used to obtain the results on particle spectra
 is the Time Projection Chamber  (TPC) \cite{TPC}.
The TPC is the primary tracking device at STAR and can track up
to 4000 charged particles per event.
It is 4.2 m long and 4 m in diameter.
Its acceptance covers $\pm 1.8$ units of pseudo-rapidity
$\eta$ and the full azimuthal angle.
The sensitive
volume of the TPC contains P10 gas (10\% methane,
90\% argon) is regulated at 2 mbar above atmospheric
pressure. The TPC data are used to determine particle
trajectories, momenta, and particle-type through
ionization energy loss (dE/dx). STAR's solenoidal
magnetic field used for this low energy Au+Au test
run was 0.5T.
In the future, the Time of Flight (ToF) detector \cite{ToF}
(with $2\pi$ azimuthal coverage and $|\eta| < 1.0)$
will further enhance the PID capability.
All events were taken with a minimum bias trigger.
The trigger detectors used in this data are the
Beam-Beam-Counter (BBC) and Vertex Position Detector (VPD)\cite{BBC}.
The BBCs are scintillator
annuli mounted around the beam pipe beyond the east
 and west pole-tips of the STAR magnet at
about 375 cm from the center of the nominal interaction
region (IR), and they have a $\eta $ coverage
of $3.8<|\eta|<5.2$ and a full azimuthal ($2\pi$) coverage.
The VPDs are based on the conventional technology
of plastic scintillator read-out by photomultiplier tubes.
They consist of two identical detector
setups very close to the beam pipe, one on each
side at a distance of $|V_z| = 5.6$~m from the center
of the IR.
The details about the design and the other characteristics
of the STAR detector can be found in \cite{TPC}.

Centrality selection for Au+Au collisions at $\sqrt {s_{NN}}$= 9.2~GeV
is defined using the uncorrected number of charged
particle tracks reconstructed in the main TPC over
the full azimuth, pseudo-rapidity $|\eta|<$~0.5 and $|V_z|<$~75~cm.
Those primary tracks which originate within 3~cm of
the primary vertex (distance of the closest approach or DCA)
and have transverse momentum $p_T>0.2$~GeV/c were selected for the analysis.
The spectra of charged hadrons were corrected for
total reconstruction efficiencies obtained
by using efficiencies and yields  of identified particles ($\pi, K,p,\bar p$)
\cite{AuAu9GeV} from embedding Monte-Carlo (MC) tracks
into real events at the raw data level
and subsequently reconstructing these events.
The background for identified particles 
was estimated in \cite{AuAu9GeV}. 
For charged hadrons it was estimated to be about $\sim 10\%$ 
at low $p_T$ and decreases up to $\sim 2\%$ at higher $p_T$.

Figure 1 shows the uncorrected multiplicity distribution
for charged tracks from the real data.
The centrality classes 0--10\%, 10--30\%, 30--60\% include
483, 1049, 1391 events
with the mean value $< N_{ch}>$ of charged tracks
$199.7\pm 1.2, 113.5\pm 0.8, 41.5\pm 0.4$, respectively.
The results presented in this paper cover the
collision centrality range of 0–-60\%. The results from
more peripheral collisions are not presented due to
large trigger inefficiencies in this test run,
which bias the data in this region \cite{AuAu9GeV}.

\section{Results and discussion}
\subsubsection{Spectra}

 The transverse momentum spectrum of
hadrons produced in high energy collisions of heavy ions
reflects features of constituent interactions
in the nuclear medium.
The medium modification is one of the effects
(recombination, coalescence, energy loss, multiple scattering,...)
 that affects the shape of the spectrum.
 The properties of the created medium
are experimentally studied by variation of the event centrality and collision energy.

Figure 2 shows the charged hadron yields in Au+Au collisions at
$\sqrt {s_{NN}} = 9.2$~GeV and mid-rapidity $|\eta|<0.5$
as a function of transverse momentum $p_T$.
The results are shown for the collision centrality
classes of 0–-10\%, 10-–30\%, 30-–60\%, and 0–-60\%.
The distributions are measured in the momentum range
 $0.2<p_T<4.$~GeV/c.
The multiplied factor of 10 is used for visibility.
As seen from Fig. 2 spectra fall more than four orders of magnitude.
The shape of the spectra indicates  the exponential and
power-law behavior at $0.2<p_T<1.$~GeV/c  and $p_T>1.$~GeV/c, respectively.
\begin{figure}
\hspace*{0mm}
\includegraphics[width=85mm,height=120mm]{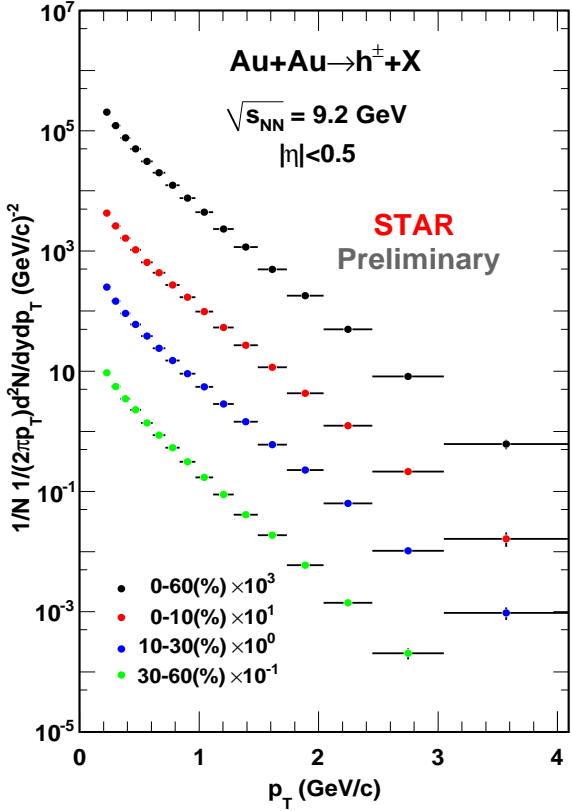}
\caption{
Mid-rapidity  ($|\eta|<  0.5$)
transverse momentum spectra for charged
hadrons produced in  Au+Au collisions
 and energy  $\sqrt {s_{NN}} = 9.2$~GeV for
various  (0--10\%, 10--30\%, 30--60\%, and 0--60\%) centralities.
The errors shown are statistical only.}
\end{figure}

The centrality dependence of $<p_T>$ is of interest,
as for a thermodynamic system
 this quantity correlates with the temperature of the system,
whereas  $dN/d\eta \propto  ln(\sqrt {s_{NN}})$
has relevance to its entropy \cite{VanHove}.
The mean values of the transverse momentum $<p_T>$
for the centrality classes 0--10\%, 10--30\%, 30--60\%,
and 0--60\% are found
to be   $553.1 \pm 1.2$~MeV/c,
 $545.4 \pm 1.1$~MeV/c,
$522.4 \pm 1.4$~MeV/c, and $543.9 \pm 0.7$~MeV/c, respectively.
The value of $<p_T>$ slowly increases with centrality.
Similar behavior is observed for pions at $\sqrt {s_{NN}}=9.2$ and 200~GeV \cite{AuAu9GeV}.
Rapid growth of $<p_T>$ vs. $dN_{ch}/d\eta$
could be associated with enhancement of multiparticle interactions
in the medium.

\subsubsection{The $R_{mult/(0-60\%)}$  and $R_{CP}$ ratios}

The ratio of transverse momentum yields for different
centralities allows us to study features of constituent
interactions in the medium depending on the scale.
Strong sensitivity  of the ratio
$R_{mult/minbias}$ ( $\simeq 0.1$ and 3.
for $dN_{ch}/d\eta =1.97$ and 9.01)
at ${p_T\simeq 4}$~GeV/c
was observed even in p+p collisions for strange particle
($K_S^0$, $\Lambda$) production \cite{pp200strange}.

Figure 3 shows the $R_{mult/(0-60\%)}$ ratio
of multiplicity binned $p_T$ spectra
to multiplicity-integrated  ($0-60\%$) spectra scaled
by the mean multiplicity in each bin for charged hadrons
\begin{equation}
R_{mult/(0-60\%)}=F_{scale}\frac{d^{2}N^{mult}/2\pi p_T dy dp_T}
{d^{2}N^{0-60\%}/2\pi p_T dy dp_T},
\label{eq:r1}
\end{equation}
where the factor $F_{scale}$ is defined as follows
\begin{equation}
F_{scale}=\frac{N_{evnt}^{0-60\%} <N_{ch}^{0-60\%}> } {N_{evnt}^{mult} <N_{ch}^{mult}> }.
\label{eq:r2}
\end{equation}
As seen from Fig. 3 the ratio is sensitive to centrality for high $p_T$.
 It increases from 0.6 to 1.2 at $p_T\simeq 3.$~GeV/c
 for low and high centralities, respectively.

Figure 4 shows the dependence of the $R_{CP}$ ratio of yields for
 the central (C) \ 0--10\%
and the peripheral (P) \ 30--60\% multiplicity classes  on the transverse momentum
\begin{equation}
R_{CP}=\frac{d^{2}N^{C}/2\pi p_T dy dp_T \   /<N_{bin}^{C}>  }
{  d^{2}N^{P}/ 2\pi p_T dy dp_T \ /<N_{bin}^{P}>}.
\label{eq:r3}
\end{equation}

Errors shown for data are the quadrature sum
of statistical uncertainties.
The ratio increases with transverse momentum.
The similar trend is observed in Au+Au collisions
at $\sqrt {s_{NN}}=200$~GeV for $p_T<1.5$~GeV/c
even though the centrality classes used for definition of $R_{CP}$ are different
\cite{AuAu200suppression}.
For $p_T>1.$~GeV/c the ratio is higher than unity,
while the $R_{CP}$ at 200 GeV never go above unity,
it decreases for $p_T>1.5$~GeV/c and
becomes approximately constant for $5<p_T< 12$~GeV/c.

\begin{figure}
\hspace*{0mm}
\includegraphics[width=77mm,height=77mm]{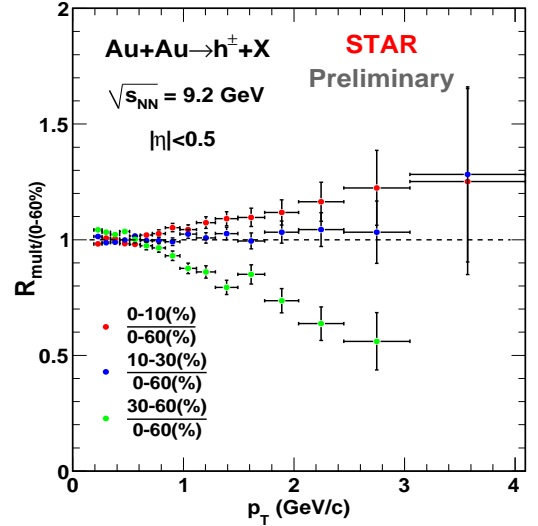}
\caption{
The $R_{mult/(0-60\%)}$  ratio of charged
hadron yields in Au+Au collisions
at mid-rapidity ($|\eta|<  0.5$) and energy  $\sqrt {s_{NN}} = 9.2$~GeV for
various  (0-10\%, 10-30\%, 30-60\%, 0-60\%) centralities
as a function of transverse momentum.
The errors shown are statistical only.}
\end{figure}

\begin{figure}
\hspace*{0mm}
\includegraphics[width=80mm,height=77mm]{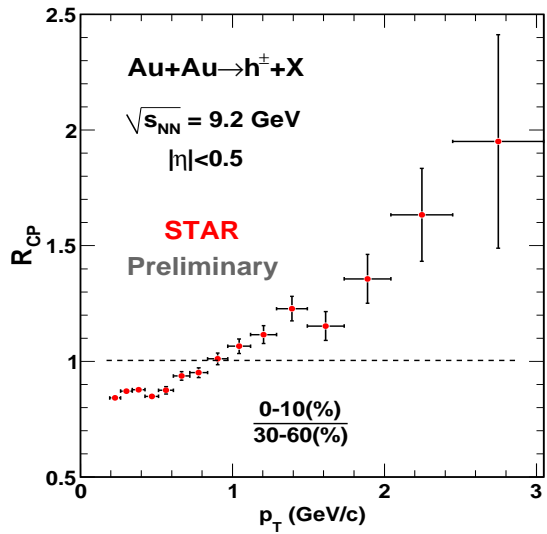}
\caption{
The $R_{CP}$  ratio of charged
hadron yields in Au+Au collisions
at mid-rapidity ($|\eta|<  0.5$) and energy
 $\sqrt {s_{NN}} = 9.2$~GeV  as a function of transverse momentum.
The errors shown are statistical only.}
\end{figure}

\subsubsection{Constituent energy loss}

The energy loss of particles created
in heavy ion collisions characterizes properties of the nuclear medium.
The nuclear modification factor $R_{AA}$ measured
at RHIC at $\sqrt {s_{NN}}=62.4,130$ and 200~GeV
strongly shows a suppression of the charged hadron spectra
at $p_T>4$~GeV/c \cite{BR1,PX1,PS1}.
These results are widely theoretically discussed
(see \cite{AuAu130suppression,AuAu200suppression} and references therein).

 The nuclear modification factor $R_{AuAu}$ for peripheral
collisions at $\sqrt {s_{NN}}=200$~GeV is close to unity at $p_T>4$~GeV/c,
while for central collisions a suppression of up to a factor
of 5 is observed.  This suppression was one of the first
indications of a strong final state
modification of particle production in Au+Au collisions
that is now generally ascribed to energy loss
of the fragmenting parton in the hot and dense medium.
The study of the evolution of the energy loss with collision
energy has relevance to the evolution of created nuclear
matter, and can be useful for searching for signature
of phase transition and a Critical Point \cite{Z}.

The measured spectra (Fig.2) allow us to estimate
constituent energy loss in charged hadron production
in Au+Au collisions at $\sqrt {s_{NN}} = 9.2$~GeV
and compare it to values obtained from
Au+Au  collisions with $\sqrt {s_{NN}} = 200$~GeV.
The  estimations are based on
a microscopic scenario of particle production  proposed in \cite{Z}.
The approach relies on a hypothesis about self-similarity
of hadron interactions at a constituent level.
The assumption of self-similarity  transforms to the requirement
of simultaneous description of transverse momentum spectra corresponding
to different collision energies, rapidities, and centralities
by the same scaling function $\psi(z)$ depending on a single variable $z$.
The scaling function is expressed in terms of the experimentally
measured inclusive invariant cross section, the multiplicity
density, and the total inelastic cross section.
It is interpreted as a probability density
to produce an inclusive particle with the corresponding value of $z$.
The scaling variable $z$ is expressed via
momentum fractions $(x_1, x_2, y_a)$,
multiplicity density, and three parameters $(\delta, \epsilon, c$).
The constituents of the incoming nuclei carry fractions $x_1, x_2$
of their momenta.
The inclusive particle carries the momentum fraction $y_a$
of the scattered constituent.
The parameters $\delta$  and  $\epsilon$ describe
 structure of the colliding nuclei
 and fragmentation process, respectively.
The parameter $c$ is interpreted
as a "specific heat" of the created medium.
Simultaneous description of different spectra with the same
$\psi(z)$ puts strong constraints on the values of these parameters,
  and thus allows for their determination.
It was found that $\delta$ and $c$ are constant and $\epsilon$ depends on multiplicity.
For the obtained values of $\delta, \epsilon$ and $c$,
the momentum fractions are determined to minimize the resolution
$\Omega^{-1}(x_1,x_2, y_a, y_b) $,
which enters in the definition of the variable $z$.
The system of the equations
$\partial \Omega/\partial x_1 = 
\partial \Omega/\partial x_2 =
\partial \Omega/\partial y_a =
\partial \Omega/\partial y_b =0$
was numerically resolved under the constraint
$(x_1P_1+x_2P_2-p/y_a)^2=(x_1 M_1+x_2 M_2+m/y_b)^2$, which
has sense of the momentum conservation law of a constituent subprocess \cite{Z}. 

The scaling  behavior of $\psi(z)$ for charged hadrons in Au+Au collisions
at ${\sqrt s_{NN}} = 200$ and 9.2~GeV
is consistent with $\delta = 0.5$ and 0.1, respectively.
The parameter $c$ was found to be 0.11.
This value is less than one (0.25) determined from $pp$ data \cite{Z}.
In this approach the energy loss of the scattered
constituent during its fragmentation in the inclusive particle
is proportional to the value $(1-y_a)$.

\begin{figure}
\hspace*{0mm}
\vskip 5mm
\includegraphics[width=78mm,height=75mm]{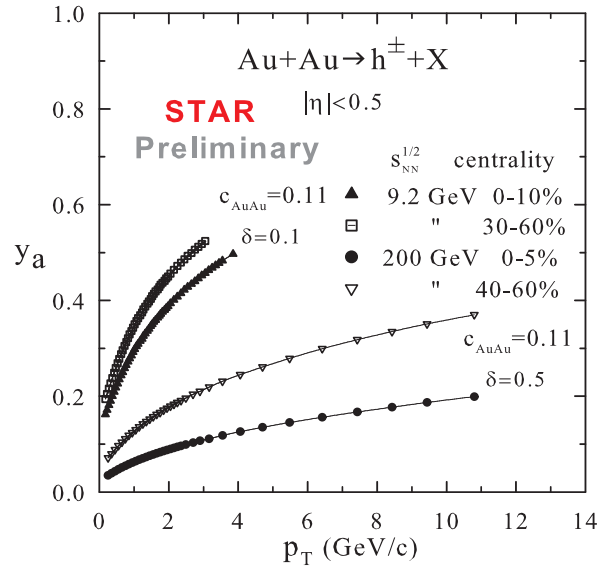}
\vskip -3mm
\caption{
The momentum fraction $y_a$ for  charged
hadron production in Au+Au collisions
at mid-rapidity ($|\eta|< 0.5$)  as a function of
the energy, centrality collision and hadron transverse momentum.}
\end{figure}

Figure 5 shows the dependence of the fraction
$y_a$ on the centrality of Au+Au collision and transverse momentum
at $\sqrt {s_{NN}}=9.2$ and 200~GeV.
The behavior of $y_a$ demonstrates a monotonic growth with $p_T$.
It means that the energy loss associated with the production of
a high-$p_T$ hadron is smaller than for hadron
with lower transverse momenta.
The decrease of $y_a$ with centrality collision
represents larger energy loss in the central
collisions as compared with peripheral interactions.
The energy dissipation grows as the collision energy increases.
It is estimated to be about 50\% at $\sqrt {s_{NN}}=9.2$~GeV and
80--90\%  at $\sqrt {s_{NN}}=200$~GeV  for $p_T\simeq 3$~GeV/c, respectively.

The saturation of hadron production established in \cite{Z}
at low $z$ (low $p_T$) is governed by the single parameter $c$.
The value of $c$ was found to be constant in Au+Au collisions
at $\sqrt {s_{NN}}= 9.2,\ 62.4,\ 130$, and 200~GeV.
Discontinuity of this parameter
was assumed to be a signature
of phase transition and a Critical Point.

\section{Summary and Outlook}

In summary,  we have presented the first STAR results for
charged hadron production in Au+Au collisions
at $\sqrt {s_{NN}}=9.2$~GeV. The spectra and  ratios of particle yields
at  mid-rapidity are measured over the range of $0.2< p_T <4.$~GeV/c.
  The centrality dependence of the hadron yields and ratios are studied.
We observed that the sensitivity of the ratios $R_{mult/(0-60\%)}$
and  $R_{CP}$ to centrality is enhanced at high $p_T$.
Hadron yields can be used to estimate of a constituent energy loss.
The energy loss of the secondary constituents passing through
the medium created in the Au+Au collisions was estimated
in the $z$-scaling approach.
It depends on the collision energy, transverse momentum, and centrality.
It was shown that the energy loss increases with the
collision energy and centrality, and decreases with $p_T$.

These results provide an additional motivation
for the Beam Energy Scan program at the RHIC \cite{BES}.
At the STAR experiment, the large and uniform acceptance
and extended particle identification (TPC, ToF) is suitable
for a Critical Point search at low energy $\sqrt {s_{NN}}=5-39$~GeV.
The study of the transition regime is interesting with respect
 to a possible modification of the
number constituent quark $v_2$-scaling, high-$p_T$ hadron suppression,
and the ridge formation, which take place in Au+Au collisions
 at higher energy at the RHIC.

\end{document}